\documentclass[mathpazo]{cicp}
\usepackage{subfigure}
\usepackage{graphics}

\begin{document}
\title[Cross Correlators & Galilean Invariance in Fluctuating LB]{Cross Correlators and Galilean Invariance in Fluctuating Ideal Gas Lattice Boltzmann Simulations}

\author[G.~Kaehler]{Goetz~Kaehler\affil{1}\corrauth and Alexander~Wagner\affil{1}}
\address{\affilnum{1}Department of Physics, North Dakota State University, Fargo, ND 58108, U.S.A.}
\emails{{goetz.kaehler@ndsu.edu} (G.~Kaehler)}


\begin{abstract}
We analyze the Lattice Boltzmann method for the simulation of fluctuating hydrodynamics by Adhikari {\em et al.} [Europhys. Lett. {\bf71}, 473 (2005)] and find that it shows excellent agreement with theory even for small wavelengths as long as a stationary system is considered. This is in contrast to other finite difference and older lattice Boltzmann implementations that show convergence only in the limit of large wavelengths. In particular cross correlators vanish to less than $0.5\%$. For larger mean velocities, however, Galilean invariance violations manifest themselves through errors of a magnitude similar to those of the earlier implementations.
\end{abstract}


\maketitle

\section{Introduction}
Fluctuations are important for many hydrodynamic phenomena, from colloid diffusion to phase-separation close to the critical point. Particle based methods such as Stochastic Rotation Dynamics~\cite{SRD}, Lattice Gas~\cite{LG} or Molecular Dynamics simulations~\cite{MD} naturally give rise to stochastic noise.
In contrast the lattice Boltzmann (LB), or finite difference discretization of the Navier Stokes equations require fluctuations that have to be included manually. 
The guiding principle for doing this is the theory of the fluctuating Navier Stokes equations \cite{ll9}. Despite the success of applying the Navier Stokes equations to very small-scale flows formally the hydrodynamic limit requires large wavelengths. For fluctuating hydrodynamics the constraint of large wavelengths becomes important and standard discretization will give results that are not in agreement with statistical physics for shorter wavelengths. For a detailed analysis of simulating fluctuating hydrodynamics using finite difference methods and some remedies to improve this situation see the recent manuscript of A.~Donev~\cite{donev1}. Similar deficiencies are found for implementations of fluctuating Navier Stokes equations using the Lattice Boltzmann approach introduced by Ladd~\cite{ladd1}. It is, however possible to use a more fundamental approach to include fluctuations in the LB method. Adhikari {\em et al.}\cite{adhikari1} introduced noise on all nonconserved modes, not only the hydrodynamic ones, leading to a scheme which shows good agreement with theory even for large wavelengths. Duenweg {\em et al.} rederived this noise implementation from detailed balance considerations of lattice gases \cite{duenweg}. Both approaches are numerically identical. In this paper we study the degree of improvement achieved and show that many of the deficiencies that plague finite difference discretizations of fluctuating Navier Stokes equations are absent in this Lattice Boltzmann implementation as long as we consider a system with vanishing mean velocity. For large mean velocities Galilean invariance is violated and errors of a similar magnitude to the earlier implementations are observed.


\section{Fluctuating Lattice Boltzmann with Ghost Noise}
Following the derivation of Adhikari {\em et al.} \cite{adhikari1} we start with the Lattice Boltzmann equation (LBE)
\begin{equation}
\label{eqn:LBE1}
f_i(\mathbf{x}+\mathbf{v}_i, t+1) = f_i(\mathbf{x}, t) + \sum_j \Lambda_{ij}\left\lbrack f_j(\mathbf{x}, t) - f_j^0(\mathbf{x},t)\right\rbrack + \xi_i(\mathbf{x}, t).
\end{equation}
Here the the $f_i$ are the particle densities at position $x$, time $t$ associated with with velocity $\mathbf{v}_i$. $\Lambda_{ij}$ is the collision matrix and $\xi_i$ are the noise terms. We use the standard local equilibrium distribution given by
\begin{equation}
\label{eqn:fnull}
f_i^0 = \rho w_i \left\lbrack 1 + \frac{3}{c^2} \mathbf{u}.\mathbf{v}_i + \frac{9}{2c^4}\left(\mathbf{u}.\mathbf{v}_i\right)^2 - \frac{3}{2c^2}\mathbf{u}.\mathbf{u}\right\rbrack,
\end{equation}
which is the discretized version of a Maxwell distribution \cite{he1997, qian1992}. In equilibrium the $f_i$ will fluctuate around this distribution. The noise terms $\xi_i$ must be chosen such that, in the case of isothermal Lattice Boltzmann (LB), the density $\rho=\sum_i f_i$ and momentum $\rho\mathbf{u} = \sum_i f_i \mathbf{v}_i$ are conserved, i.e. $\sum_i \xi_i = 0$ and $\sum_i \xi_i v_i = 0$. Furthermore a proper fluctuation dissipation theorem (FDT) corresponding to the collision operator $\Lambda_{ij}$ is obeyed. This implies that the $\xi_i$ are correlated. We can find a representation in which the noise terms are uncorrelated by transforming the LBE into moment space. The moments are given by
\begin{equation}
\label{eqn:transform1}
M^a(\mathbf{x}, t) = \sum_i m_i^a f_i(\mathbf{x}, t).
\end{equation}
So far this is a standard Multi-Relaxation-Time (MRT) representation \cite{dhumieres, lallemand1, adhikari2}. The back transform is given by $f_i(\mathbf{x}, t) = \sum_a n_i^a M^a(\mathbf{x}, t)$. However, in order to construct a proper FDT these transforms cannot be orthogonal as in other MRT methods~\cite{dhumieres, lallemand1}, so here we have $n_i^a \ne m_i^a$. Instead the transforms are chosen such that
\begin{equation}
\label{eqn:hermitenorm}
\sum_i w_i m_i^a m_i^b = \sum_i m_i^a n_i^b = \delta^{ab}
\end{equation}
with $n_i^a = w_i m_i^a$ while maintaining a diagonal moment space representation of the collision operator $\Lambda_{ij} =- \sum_a\sum_b n_i^a \frac{1}{\tau^a} \delta^{ab} m_j^b$. 
Now the moment transformation matrices are orthogonal with respect to the Hermite norm. Such transforms with weighted norms were proposed before \cite{vergassola1990, dellar2002, dellar2003} in different contexts. The necessity of the Hermite norm is briefly outlined after Eq.~(\ref{eqn:decouplemoments}) below and allows for a convenient definition of the moment space noise terms $\xi^a$ as independent random variables. We can now rewrite the collision term of the Lattice Boltzmann equation in terms of the moments $M^a$ as
\begin{equation}
\label{eqn:collision}
f_i(\mathbf{x}+\mathbf{v}_i, t+1) = \sum_a n_i^a \left\lbrace M^a(\mathbf{x}, t) - \frac{1}{\tau^a} \left\lbrack M^a(\mathbf{x}, t) - M^{a, 0}(\mathbf{x}, t) \right\rbrack + \xi^a \right\rbrace.
\end{equation}
Adhikari {\em et al.}\cite{adhikari1} then obtain the FDT by performing a Fourier transform of the fluctuations from the mean of the moments $\delta M^a = M^a - \langle M^a \rangle$. They then use the $k$-independence of these for an ideal gas to obtain
\begin{equation}
\label{eqn:FDT}
 \left\langle \xi^a \xi^c \right\rangle = \frac{\tau^a + \tau^c - 1}{\tau^a \tau^c} \left\langle \delta M^a \delta M^c \right\rangle.
\end{equation}
One particular result of the derivation is that the moment fluctuations $\xi^a$ decouple because
\begin{eqnarray}
\left\langle\delta M^a \delta M^b\right\rangle & = & \sum_i \sum_j m_i^a m_j^b \langle \delta f_i \delta f_j \rangle \nonumber \\ 
& = & \sum_i \sum_j m_i^a m_j^b \bar{f}_i \delta_{ij}\nonumber \\
& = & \sum_i m_i^a m_i^b \bar{\rho} w_i \nonumber \\
\label{eqn:decouplemoments}
& = & \bar{\rho} \delta^{ab} .
\end{eqnarray}
Here we used $\langle \delta f_i \delta f_j \rangle = \bar{f}_i \delta_{ij}$ with $\delta f_i = f_i - \bar{f}_i$ where $\bar{f}_i$ is the spatially uniform global equilibrium distribution function \cite{ll10}. Adhikari also assumed that $\mathbf{u} \ll 1$ so that $\bar{f}_i = \bar{\rho} w_i$. This allows us to use the orthogonality relation of Eq.~(\ref{eqn:hermitenorm}) in the last step of the calculation above. For a different transformation we would obtain non-diagonal elements in the fluctuation matrix which will then require correlated noise terms which are more cumbersome to implement. For practical applications it is important to note that the $\mathbf{u} \ll 1$ condition for the noise introduces a non-Galilean invariant contribution. We comment on this in our validation section.
Inserting Eq.~(\ref{eqn:decouplemoments}) into Eq.~(\ref{eqn:FDT}) leads to a noise expression of 
\begin{equation}
\label{eqn:noiseamp}
\xi^a = \frac{1}{\tau^a} \sqrt{\bar{\rho} \left(2 \tau^a - 1\right)}N,
\end{equation}
where $N$ is a random variable with zero mean and a variance of one. 

Note that the moments $M^a$ are chosen to include the hydrodynamic moments. In the isothermal case discussed here they are comprised of the conserved quantities $\rho$ and $\mathbf{j}$, and the stress modes $\mathbf{\Pi}$.  The remaining degrees of freedom are often called ghost modes as they do not appear in the isothermal Navier Stokes equations. However, the key result of the Adhikari {\em et al.} paper \cite{adhikari1} was that they need to be taken into account when including noise. Thus we add noise on all non conserved quantities, i.e. stress and ghost modes, in Eq.~(\ref{eqn:collision}) according to Eq.~(\ref{eqn:noiseamp}). 

In practice we implement this algorithm by calculating the moments by means of Eq.~(\ref{eqn:transform1}), performing the collision on the moments, adding the noise term and then transforming back into $f$-space as indicated in Eq.~(\ref{eqn:collision}). The streaming step is then done in $f$-space. This algorithm is almost as efficient as the standard LB implementation. The additional computational cost for calculating the ghost modes and the random numbers results in a computational overhead of less than $20\%$.

\section{Correlators in a D2Q9 Implementation}

\begin{figure}
\subfigure[Without ghost noise]{\includegraphics[scale = .4]{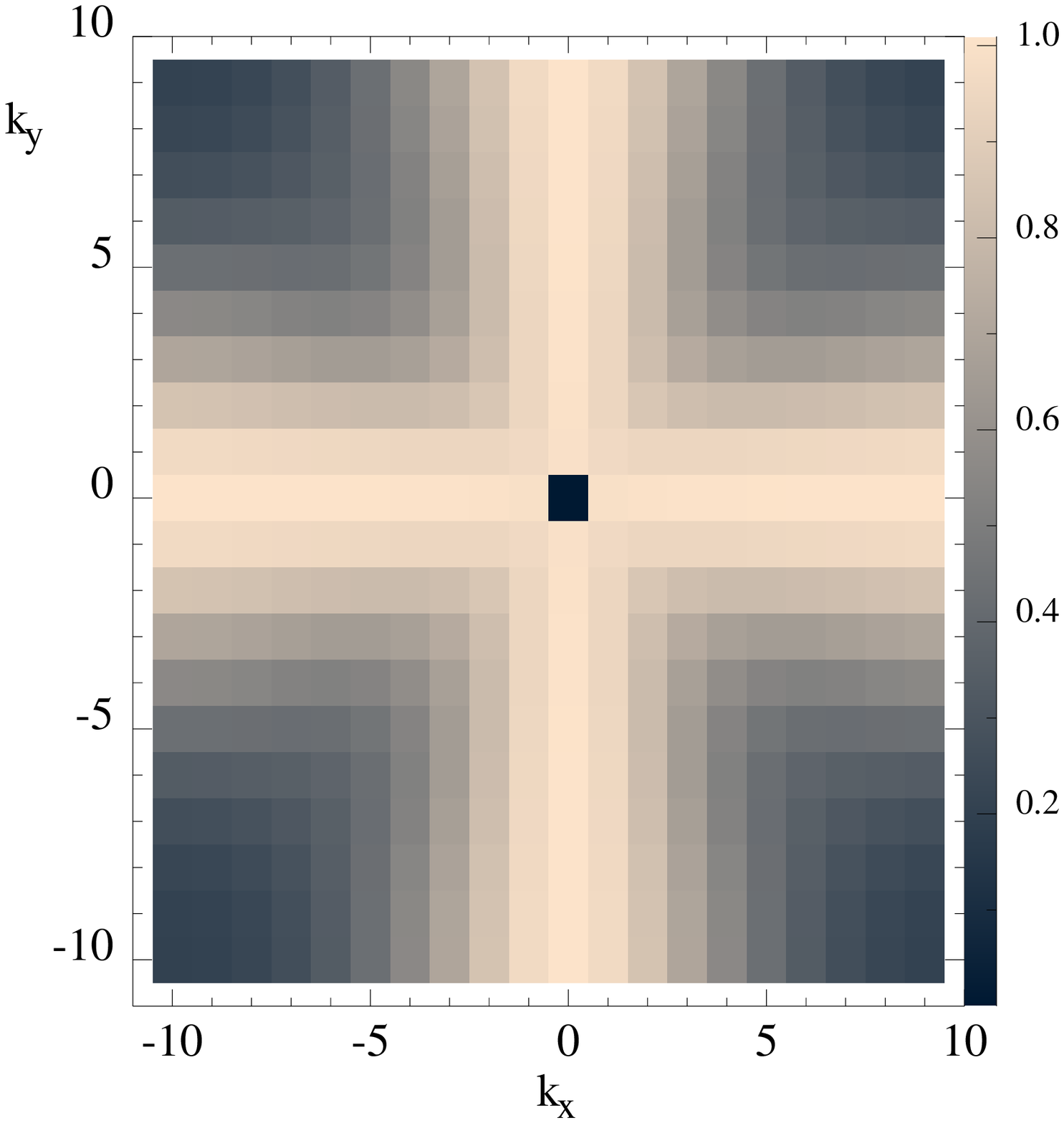}}
\subfigure[With ghost noise]{\includegraphics[scale = .4]{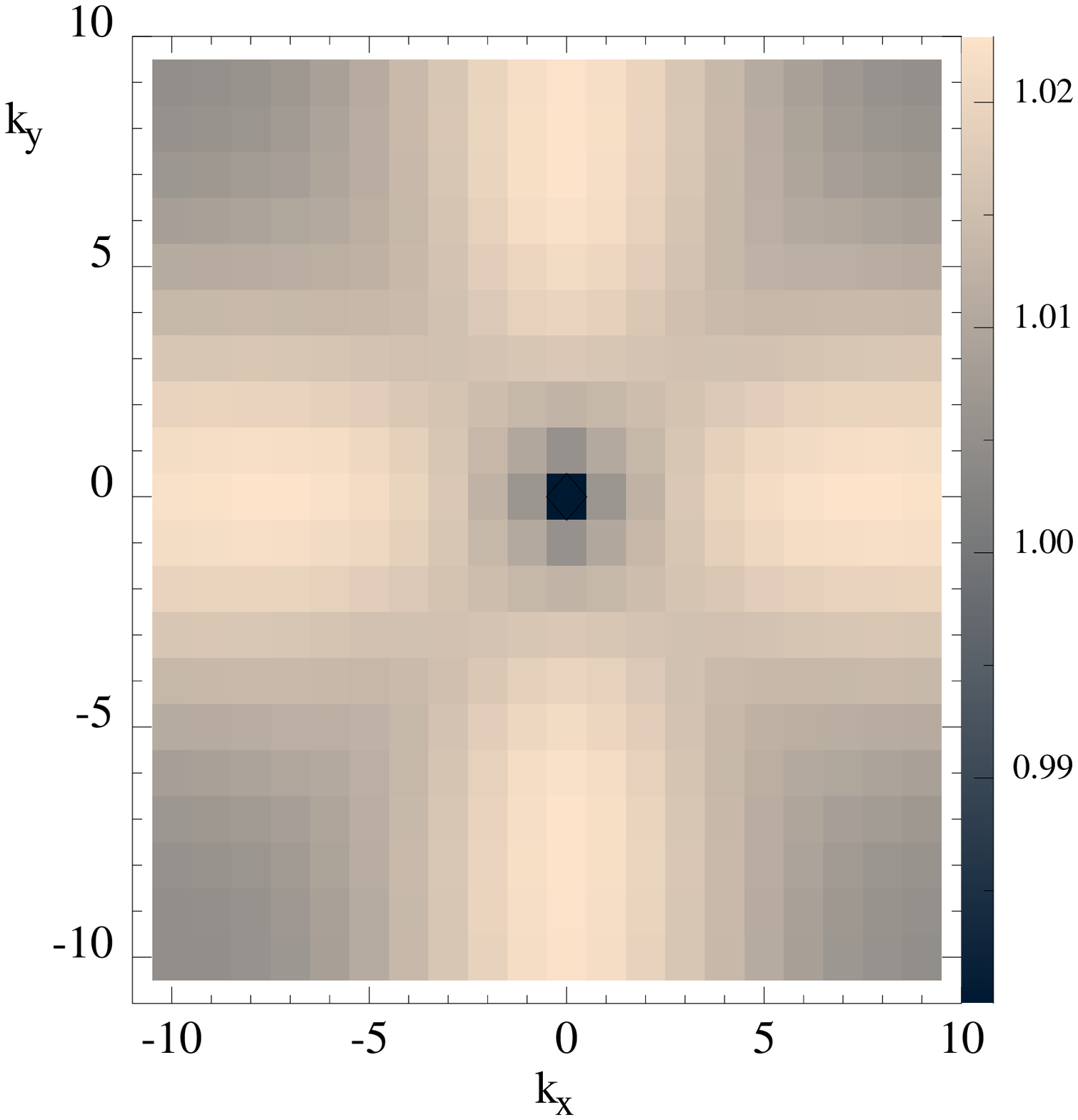}}
\caption{$S_{\mathbf{k}}(\rho)$ averaged over $2\times10^8$ iterations in a  $\tau^a=1$ for all $a$, $V=20^2$, fluctuating D2Q9 ideal gas without and with active ghost noise. Note that different scales are used to visualize the slight deviations seen in the ghost noise case.}
\label{fig:d2q9grr}
\end{figure}

\begin{figure}
\subfigure[Without ghost noise]{\includegraphics[scale = .4]{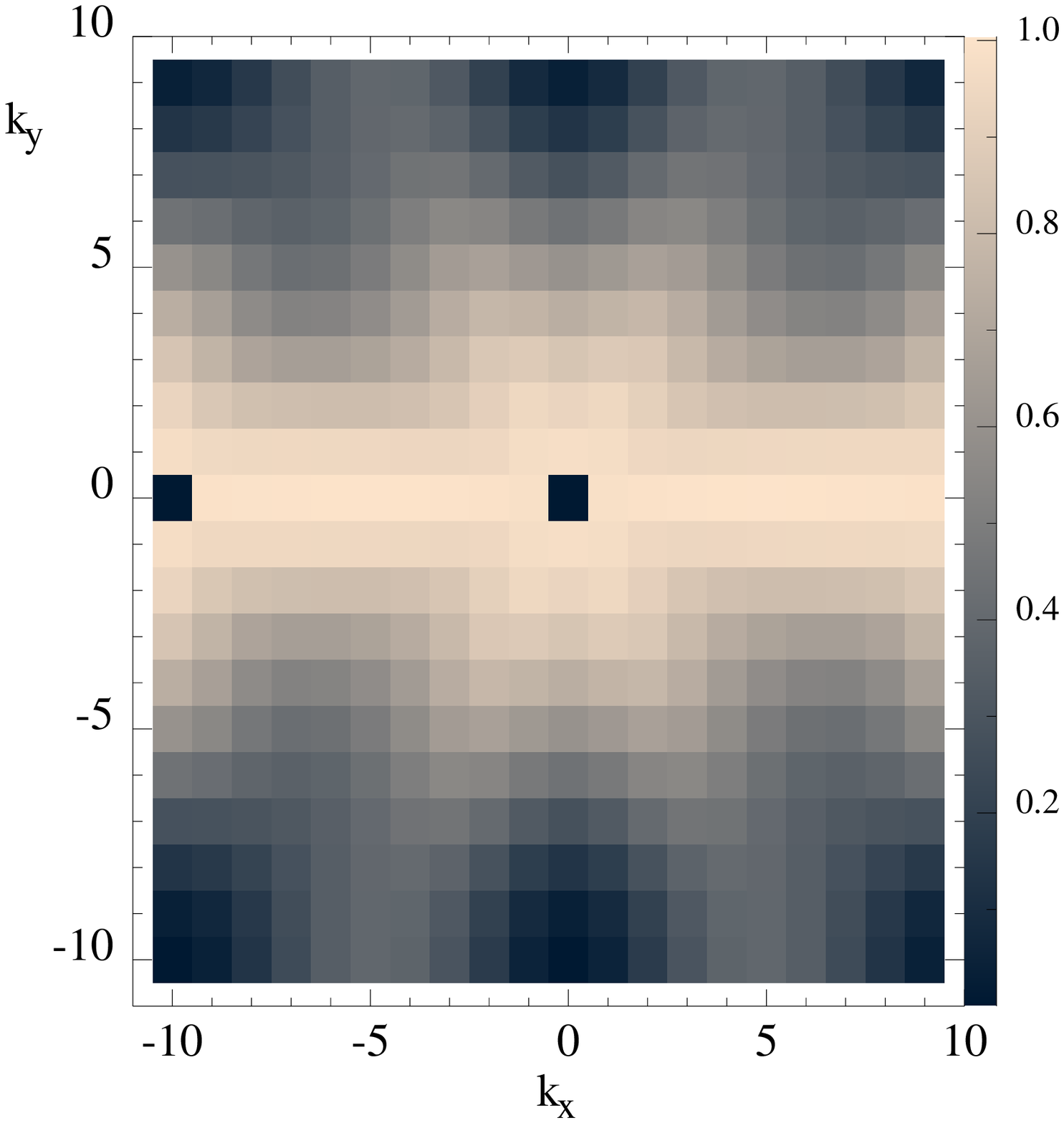}}
\subfigure[With ghost noise]{\includegraphics[scale = .4]{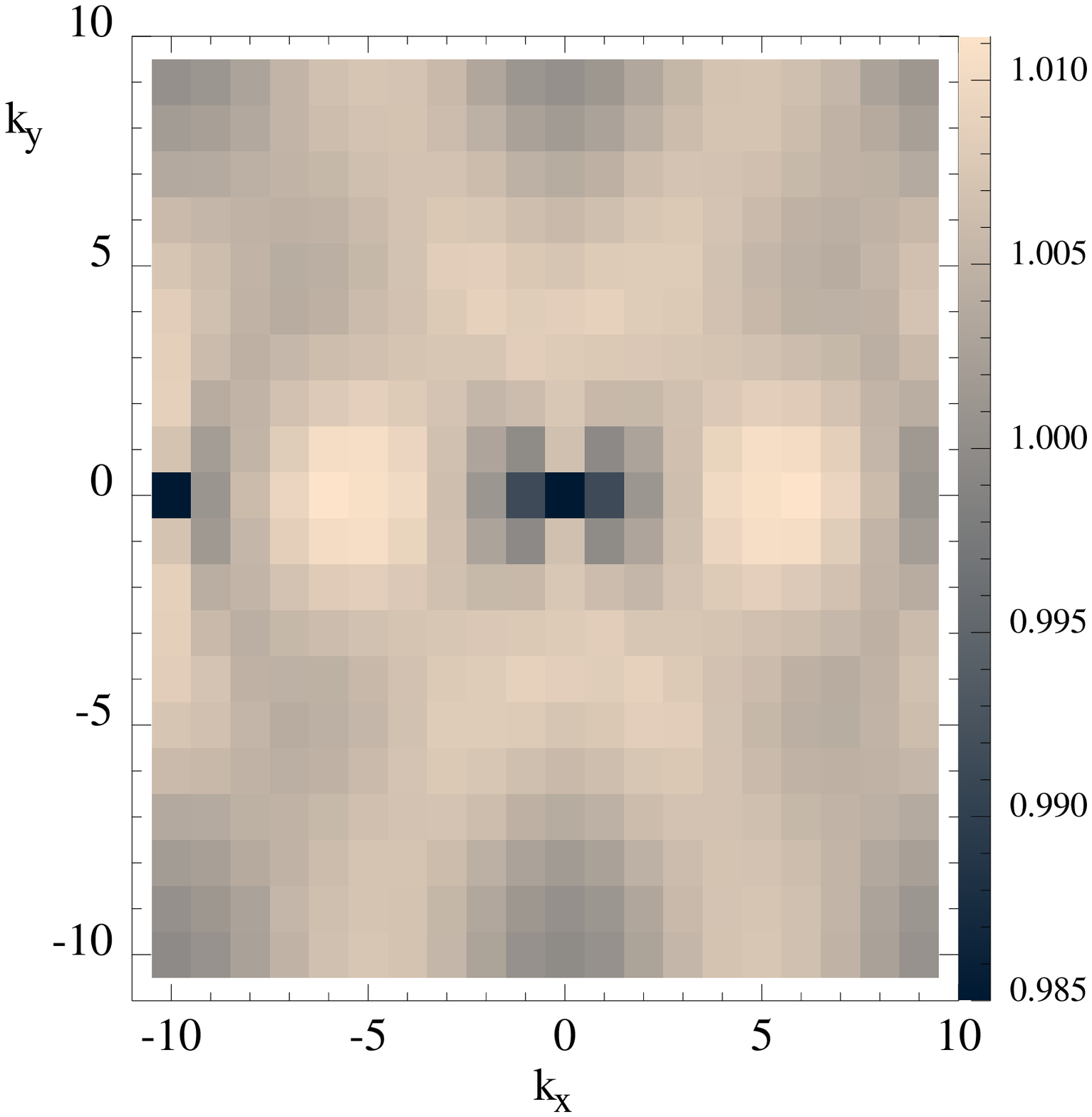}}
\caption{$S_{\mathbf{k}}(u_x)$ averaged over $2\times10^7$ iterations in a  $\tau^a=1$ for all $a$, $V=20^2$, fluctuating D2Q9 ideal gas without and with active ghost noise. Note that different scales are used to visualize the slight deviations seen in the ghost noise case.}
\label{fig:d2q9guxux}
\end{figure}

To evaluate this method we present results for the D2Q9 (two dimensions, 9 base velocity vectors) LB model. The results are similar for other models, in particular we also tested D1Q3 and D3Q15. As D2Q9 base velocity set we use $\{v_i\} = \{(0,0), (1,0), (0,1),$ $(-1,0), (0,-1),$ $(1,1), (-1,1),$ $(-1,-1), (1,-1)\}$ and the $\{w_i\} = \{4/9,$ $1/9,$ $1/9,$ $1/9,$ $1/36,$ $1/36,$ $1/36,$ $1/36\}$
.
The matrix elements $m_i^a$ in transform~(\ref{eqn:transform1}) are then given by
\begin{equation}
\label{eqn:d2q93a}
\{m_i^a\} = \left(
\begin{array}{ccccccccc}
 1 &  1 &  1 &  1 &  1 &  1 &  1 &  1 &  1 \\
 0 &  \sqrt{3} &  0 &  -\sqrt{3} &  0 &  \sqrt{3} &  -\sqrt{3} &  -\sqrt{3} &  \sqrt{3} \\
 0 &  0 &  \sqrt{3} &  0 &  -\sqrt{3} &  \sqrt{3} &  \sqrt{3} &  -\sqrt{3} &  -\sqrt{3} \\
 0 &  \frac{3}{2} &  \frac{-3}{2} &  \frac{3}{2} &  \frac{-3}{2} &  0 &  0 &  0 &  0 \\
 0 &  0 &  0 &  0 &  0 &  3 &  -3 &  3 &  -3 \\
 -1 &  \frac{1}{2} &  \frac{1}{2} &  \frac{1}{2} &  \frac{1}{2} &  2 &  2 &  2 &  2 \\
 0 &  -\sqrt{\frac{3}{2}} &  0 &  \sqrt{\frac{3}{2}} &  0 &  \sqrt{6} &  -\sqrt{6} &  -\sqrt{6} &  \sqrt{6} \\
 0 &  0 &  -\sqrt{\frac{3}{2}} &  0 &  \sqrt{\frac{3}{2}} &  \sqrt{6} &  \sqrt{6} &  -\sqrt{6} &  -\sqrt{6} \\
 \frac{1}{2} &  -1 &  -1 &  -1 &  -1 &  2 &  2 &  2 &  2
\end{array}
\right).
\end{equation}
The corresponding elements $n_i^a$ of the back transform are defined by the requirement $n_i^a = m_i^a w_i$.
The zeroth moment then is the density $\rho$, the first and second are (up to a factor) the components of the momentum, the third and fourth the components of the shear stress and the fifth resembles the bulk stress \cite{kaehler2010}. The remaining three moments are the ghost modes. Thus the equilibrium moments $M^{a,0} = \sum_i m_i^a f_i^0$ are
\begin{equation}
\label{eqn:d2q94}
\begin{array}{ccc}
M^{0,0} & = & \rho\\
M^{1,0} & = &  \sqrt{3}\rho u_x \\
M^{2,0} & = &  \sqrt{3}\rho u_y \\
M^{3,0} & = &  \frac{3}{2} \rho (u_x^2 - u_y^2) \\
M^{4,0} & = &  3 \rho u_x u_y \\
M^{5,0} & = &  \frac{3}{2} \rho (u_x^2 + u_y^2) \\
M^{6,0} & = & M^{7,0} = M^{8,0} =  0 .
\end{array}
\end{equation}

We present here results for $k$-independence of the moment fluctuations predicted by Eq.~(\ref{eqn:decouplemoments}). In particular we consider the normalized static structure factor
\begin{equation}
\label{eqn:scorrelator}
S_{\mathbf{k}}(M^a) = N^a \left\langle \delta M^a(\mathbf{k}) \delta M^a(-\mathbf{k}) \right\rangle
\end{equation}
where $\delta M^a(\mathbf{k}) = \sum_{\mathbf{x}} \delta M^a(\mathbf{x}) e^{-i \mathbf{k} \cdot \mathbf{x}}$ is the discrete spatial Fourier transform of $\delta M^a$ and $\sum_{\mathbf{x}}$ is understood to be the summation over all discrete lattice sites. The normalization constant $N^a$ such that $S_{\mathbf{k}}(M^a) = 1$ is equivalent to $\bar{\rho}$. I. e. for the density $N^\rho = \frac{1}{\bar{\rho}^{3} V}$ and velocity components $N^{u_\alpha} = \frac{1}{\bar{\rho} V k_b T }$ where $k_b T = \frac{1}{3}$ for the isothermal D2Q9 model employed. A value of $1$ throughout $k$-space for the structure factor of any of the moments given in Eq.~({\ref{eqn:d2q94}}) thus indicates agreement with Eq.~(\ref{eqn:decouplemoments}). The volume $V$ is just the number of lattice points $V = \sum_{\mathbf{x}} 1$ and the division by it is just a normalization artifact of the Fourier transform. 

According to the argument put forth in \cite{adhikari1} we expect the mean square fluctuations of all moments $M^a$ to be unity throughout $k$-space. For the density $\rho$ this is confirmed to three orders of magnitude in Fig.~\ref{fig:d2q9grr}(b) for $S_{\mathbf{k}}(\rho)$ and in Fig.~\ref{fig:d2q9guxux}(b) for $S_{\mathbf{k}}(u_x)$. We find similar agreement for all nine moments of the D2Q9 model. For comparison we set the noise on the non-hydrodynamic modes ($M^6, M^7, M^8$) to zero, recovering the original Ladd method~\cite{ladd1} and, as seen in Figures~\ref{fig:d2q9grr}(a) and~\ref{fig:d2q9guxux}(a), the lack of noise on the ghost terms leads to drastic deficiencies for large $k_x, k_y$ values. Note that there are no deficiencies in Fig.~\ref{fig:d2q9grr} for $k_x = 0$ and $k_y=0$. The reason is that the projection of the D2Q9 model onto one coordinate axis yields a D1Q3 model. The isothermal ideal gas D1Q3 model, however, only has one stress mode and no ghost modes and thus there is no difference between the Ladd and Adhikari implementations in these projections. This is again observed in Fig.~\ref{fig:d2q9guxux}(a) where $S_{\mathbf{k}}(u_x)$ exhibits white noise along the $k_x$ axis even in the absence of ghost noise.
\begin{figure}
\subfigure[Without ghost noise]{\includegraphics[scale = .40]{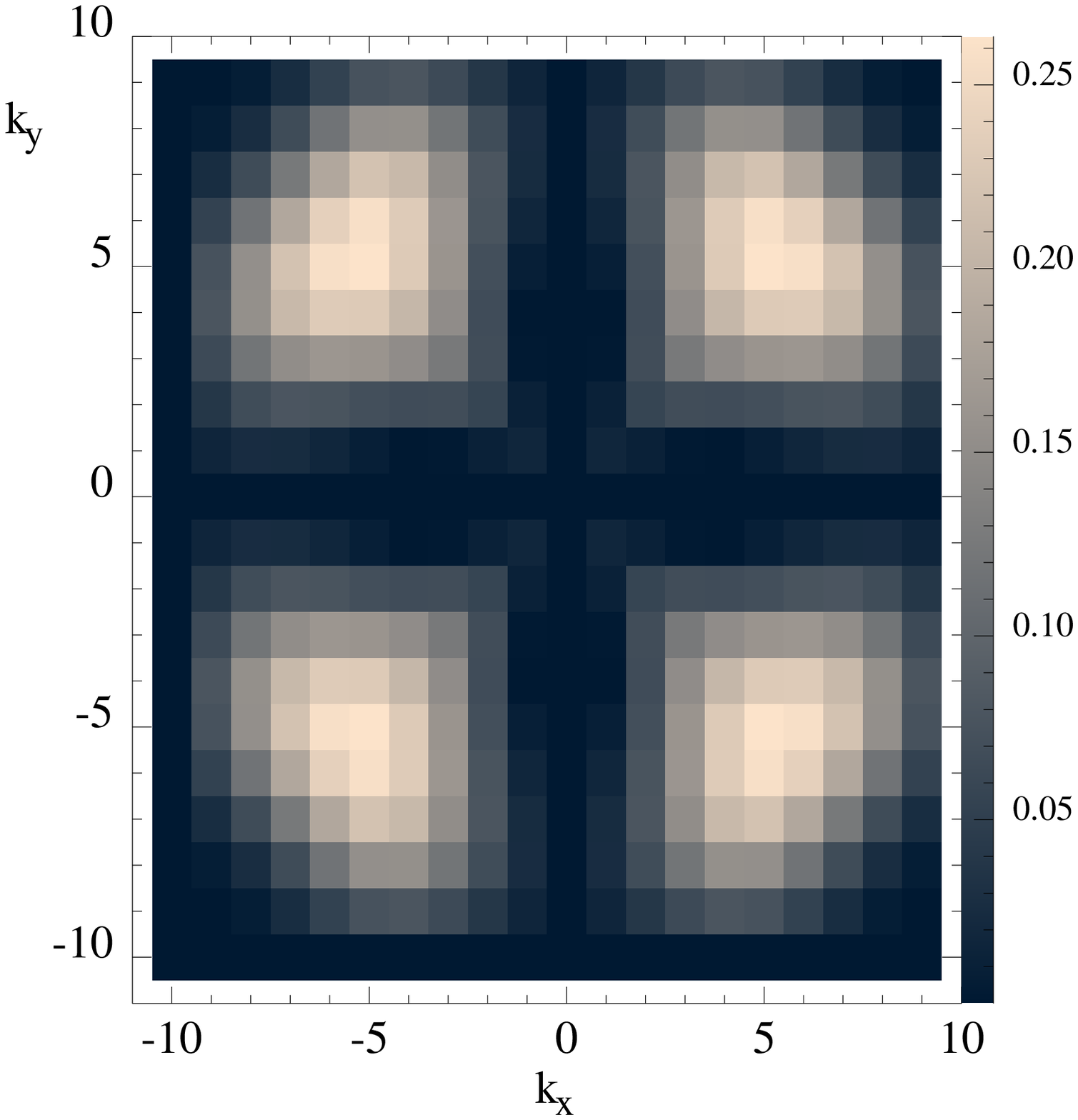}}
\subfigure[With ghost noise]{\includegraphics[scale = .40]{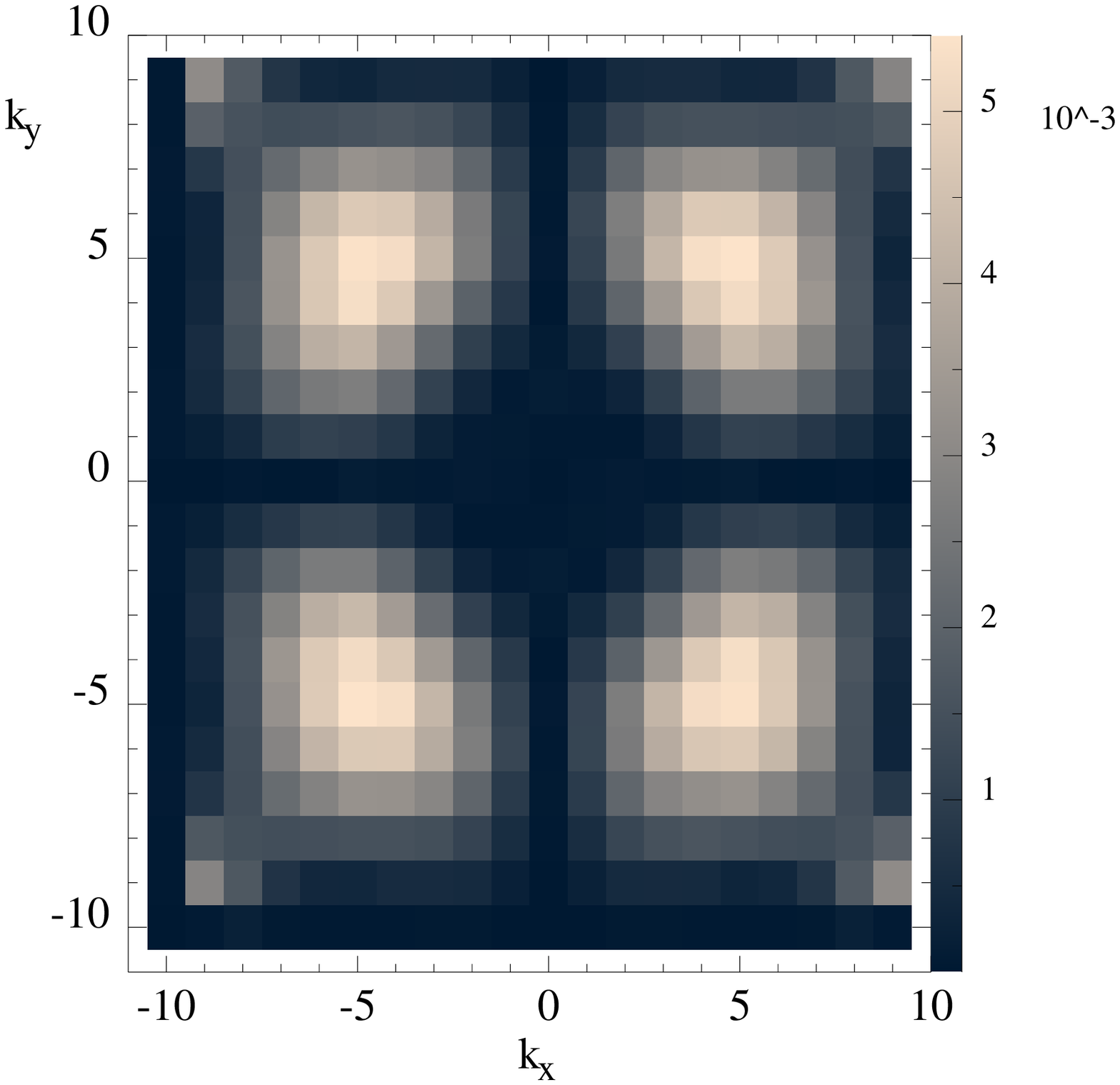}}
\caption{$R_{\mathbf{k}}(u_x u_y)$ averaged over $8\times10^6$ iterations in a  $\tau^a=1$ for all $a$, $V=20^2$ fluctuating D2Q9 ideal gas simulation with and without active ghost noise. Again, take note of the different scales.}
\label{fig:d2q9gvxvy}
\end{figure}
Motivated by private communication with A.~Donev who is developing a general finite volume scheme to solve the fluctuating Navier Stokes Equations~\cite{donev1} based on a third order Runge-Kutta integrator we also measured the cross correlator
\begin{equation}
\label{eqn:rcorrelator}
R_{\mathbf{k}}(u_x, u_y) = N^{u_x} \left\langle u_x(\mathbf{k}) u_y^*(\mathbf{k}) \right\rangle.
\end{equation}
According to Eq.~(\ref{eqn:decouplemoments}) this quantity is expected to vanish. This is again confirmed nicely in Fig.~\ref{fig:d2q9gvxvy}(b) to three orders of magnitude. In contrast measurements of $R_{\mathbf{k}}(u_x, u_y)$ in an implementation without ghost noise exhibits significant correlations of up to $0.25 \bar{\rho}$ for intermediate $k_x$ and $k_y$ ranges as seen in Fig.~\ref{fig:d2q9gvxvy}(a).

\begin{center}
\begin{figure}
\subfigure[$S_{\mathbf{k}}(\rho)$]{\includegraphics[scale = .258]{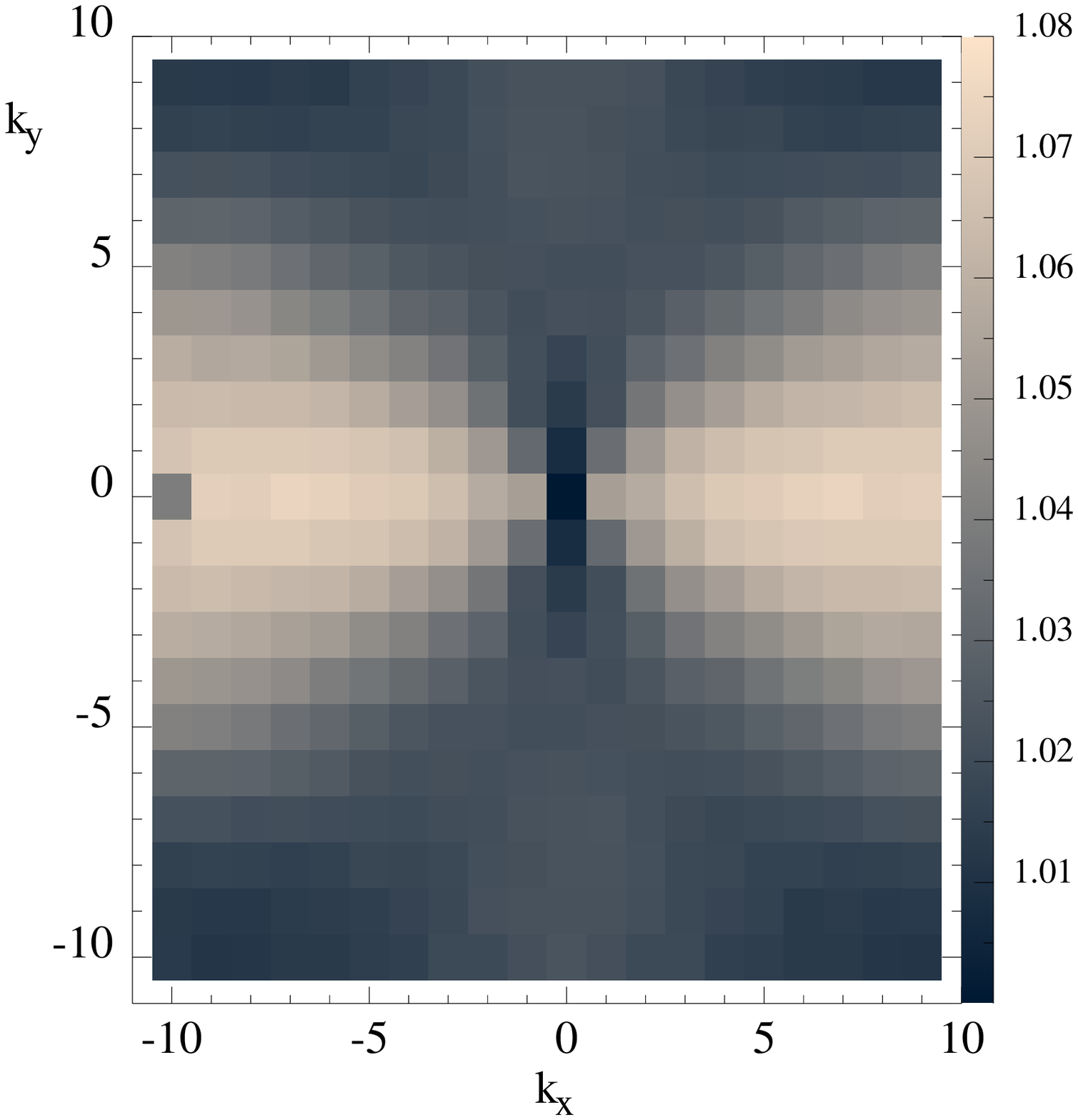}}
\subfigure[$S_{\mathbf{k}}(u_x)$]{\includegraphics[scale = .258]{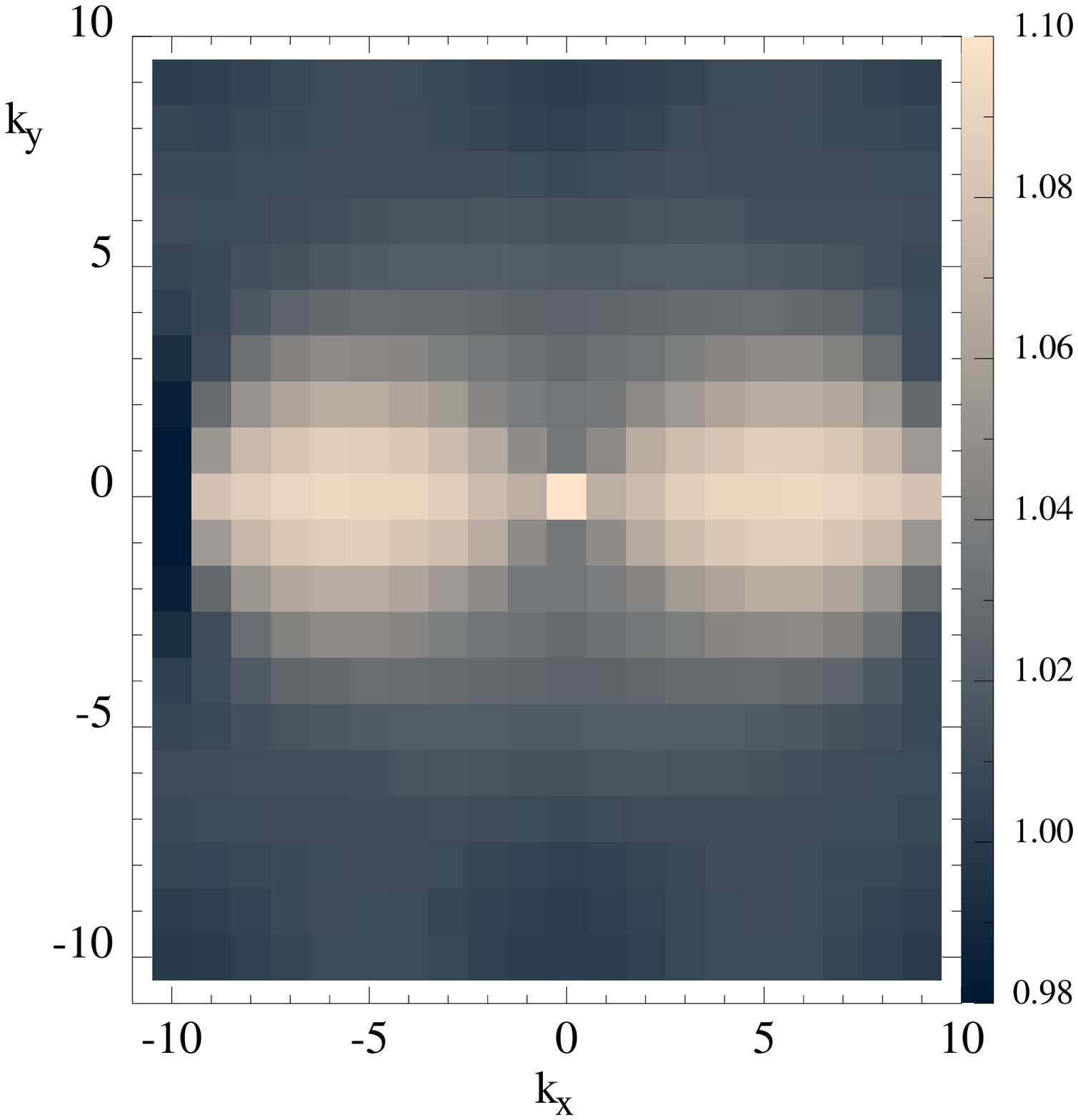}}
\subfigure[$S_{\mathbf{k}}(u_x, u_y)$]{\includegraphics[scale = .258]{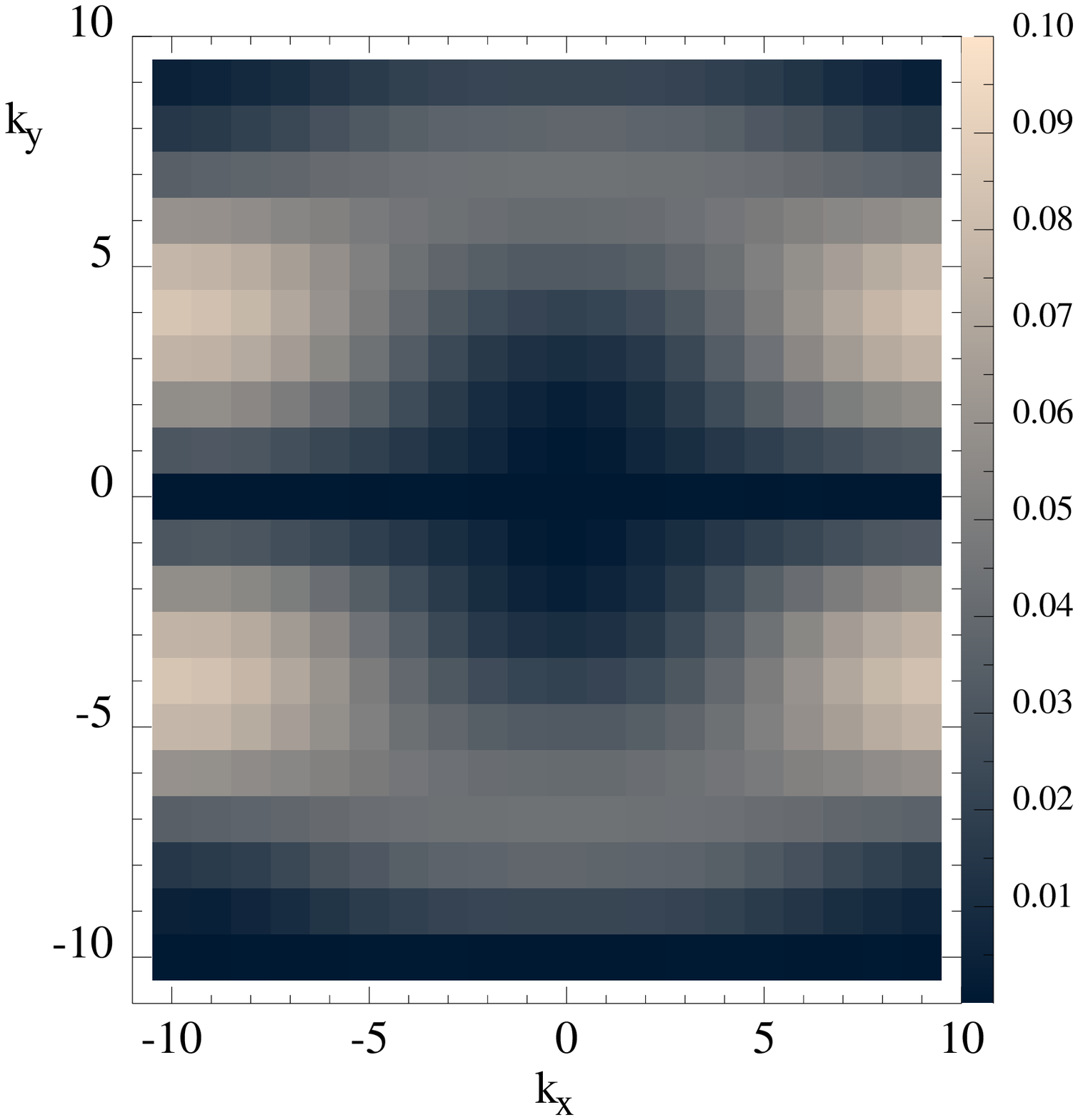}}
\caption{Correlators $S_{\mathbf{k}}(\rho)$, $S_{\mathbf{k}}(u_x)$, and $R_{\mathbf{k}}(u_x, u_y)$ averaged over $5\times10^6$ iterations or a $\tau^a=1$ for all $a$, $V=20^2$ fluctuating D2Q9 ideal gas simulation with a constant velocity of $u_x=0.1$.}
\label{fig:d2q9uxlarge}
\end{figure}
\end{center}

The required condition in Eq.~(\ref{eqn:decouplemoments}), $\mathbf{u} \ll 1$, suggests that this noise implementation may suffer from a lack of Galilean invariance. To estimate the magnitude of this violation we consider an imposed mean velocity in the $x$-direction. We measured correlators for a fluctuating system with large superimposed velocity of $u_x = 0.1$. The results in Fig.~\ref{fig:d2q9uxlarge} indicate that indeed the moment fluctuations do not decouple and Eq.~(\ref{eqn:decouplemoments}) is no longer fulfilled. Compared to the Ladd implementation these errors are still smaller, but they approach the same order of magnitude for maximal accessible velocities. A more comprehensive investigation of these effects is subject of a forthcoming publication. 

\section{Discussion and Outlook}
We have shown here that the Adhikari approach to use an improved LB method presents a promising scheme to simulate fluctuating hydrodynamics. The ability to interprete the ghost degrees of freedom as resulting from discrete particle distributions gives us the ability to systematically introduce fluctuations. This approach recovers fluctuations not only in the hydrodynamic limit but also for much shorter wavelengths. However, this is only true in the absence of flow. Since lattice Boltzmann methods are not generally used in this regime one may wonder if Galilean invariance violations may not erase much of the improvement achieved by including noise in the ghost modes. This is a subject to which we will return in a forthcoming paper.






\end{document}